\documentclass[aps,twocolumn]{revtex4-1}
\usepackage{amsfonts}
\usepackage{amsmath}
\usepackage{graphicx}
\usepackage{color}

\begin{document}

\title{Topological quantized edge-pumping-spin flip in Rice-Mele model with
spin-orbit coupling}
\author{E. S. Ma}
\author{Z. Song}
\email{songtc@nankai.edu.cn}
\affiliation{School of Physics, Nankai University, Tianjin 300071, China}

\begin{abstract}
The quantized Thouless pumping charge in a spinless Rice-Mele (RM) model
originates from a degeneracy point in the parameter space and cannot be
detected when open boundary conditions are applied. In this work, we
investigate the topological features of a spinful Rice-Mele (RM) model. We
demonstrate that spin-orbit coupling facilitates the transition of a single
degenerate point into a degenerate loop, which is anticipated to be the
source of the topological characteristics. When periodic boundary conditions
are considered, we find that the pumping spin is zero for an adiabatic loop
within the nodal loop and is 2 (in units of $\hbar /2$) for an adiabatic
passage enclosing the nodal loop. When open boundary conditions are
considered, the boundary-bulk correspondence is demonstrated by quantized
pumping-spin flips at the edges, which can be obtained by completing double
periods of a closed passage, rather than a single cycle. Our findings reveal
an alternative dynamic manifestation of the boundary-bulk correspondence.
\end{abstract}

\maketitle

\section{Introduction}

Thouless pumping, as the quantum-version of matter pumping \cite%
{Thouless1983Quan} by a mechanical device, has garnered significant
attention over a long period. It involves the transport of charge without a
net external electric or magnetic field, achieved through an adiabatic
cyclic evolution of the underlying Hamiltonian. In contrast to transport by
a classical device, the transported charge in a Thouless pump exhibits two%
\textbf{\ }intriguing features. Firstly, the total probability of the
transferred particles is precisely quantized during a cyclic adiabatic
passage. Secondly, as a demonstration of topological invariant, a nonzero
pumping charge for the ground state is shown to be related to a degenerate
point \cite{XiaoDi}. Recently, the versatility and control of synthetic
quantum systems have made experimental realization of quantum pumping
possible. Electron pumping experiments have been performed in various
semiconductor-based nanoscale devices \cite%
{Switkes1999a,Blumenthal2007,kaestner2008single}. More recently, the
topological charge pump was realized in optical superlattices based on
ultracold atom technology \cite%
{nakajima2016topological,lohse2016thouless,lu2016geometrical}, and it has
been also extensively studied in theory \cite%
{chiang1998quantum,qian2011quantum,wang2013topological,matsuda2014topological,mei2014topological,wei2015anomalous,yang2018continuously,Matsuda2022,zhang2020top}%
. {\ To date, there have been many works focusing on topological pumping
charge in different systems, such as in superconducting circuits \cite%
{Geerligs1991,Pekola1999adiabatic,fazio2003measurement, Leone2008}, in
multiterminal Josephson junctions of conventional superconductors \cite%
{Riwar2016,Xie2018} and in topological superconductors \cite%
{Teo2010,Keselman2013,Kotetes2019,Houzet2019,Mercaldo2019}. }

In this work, we studied the topological Thouless pumping of a spinful
Rice-Mele (RM) chain with spin-orbit coupling. In general, the quantized
Thouless pumping charge in a spinless RM model originates from a degeneracy
point in the parameter space and cannot be detected when open boundary
conditions are applied. We demonstrate that spin-orbit coupling facilitates
the transition of a single degenerate point into a degenerate loop, which is
anticipated to be the source of the topological characteristics. When
periodic boundary conditions are considered, analytical and numerical
studies show that the pumping charge remains zero as the degenerate point
transitions to a nodal circle. In contrast, we find that the pumping spin is
zero for an adiabatic loop within the nodal loop and is $2$ (in units of $%
\hbar /2$) for an adiabatic passage enclosing the nodal loop. In addition,
we also investigate a dynamic manifestation of the boundary-bulk
correspondence. We calculate the pumping-spin flip for different adiabatic
passages\ under the open boundary conditions. We find that the quantized
pumping-spin flips at the edges can be obtained only when double periods of
a closed passage is considered. The pumping-spin flips at the edges is zero
for an adiabatic loop within the nodal loop and is $2$ (in units of $\hbar
/2 $) for an adiabatic passage enclosing the nodal loop. Our findings not
only propose a new origin for the topology but also provide another method
for dynamically detecting the boundary-bulk correspondence.

\begin{figure*}[tbh]
\centering
\includegraphics[width=0.9\textwidth]{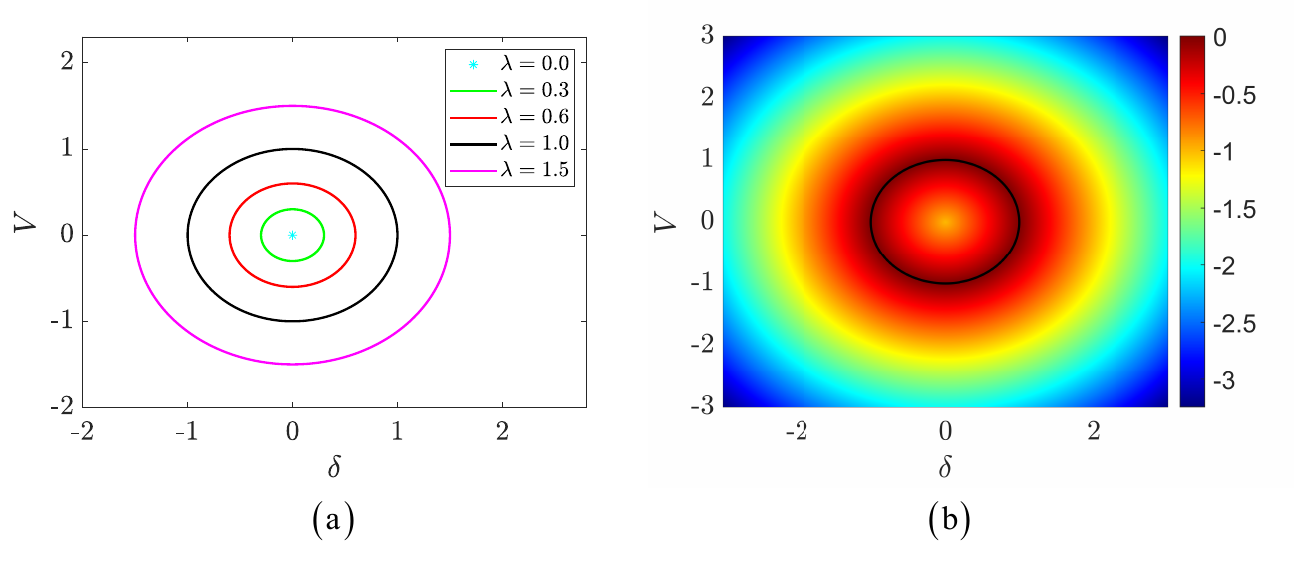}
\caption{The spectral structure in the parameter space for the Hamiltonian
given by Eq. (\protect\ref{H}). (a) Several representative nodal circles
derived from Eq. (\protect\ref{nodal circle}) on the $\protect\delta -V$
plane for different spin-orbit coupling strengths, as indicated in the
panel. (b) Color contour plots of the energy band $\protect\varepsilon %
_{-,-} $ from Eq. (\protect\ref{spectrum}), with the wavevector $k$ set to $%
\protect\pi $. The black circle represents $\protect\lambda =1$, which
corresponds to the one plotted in (a).}
\label{fig1}
\end{figure*}
The structure of the paper is outlined as follows. We commence in Sec. \ref%
{Model and symmetry} by presenting the Hamiltonian and examining its
symmetry. In Sec. \ref{Topological pumping spin}, for a straightforward
scenario where spin-orbit coupling is absent, we assess the validity of
employing pumping charge as a topological indicator. Consequently, we
introduce the concept of topological pumping spin. In Sec. \ref{Nodal circle}%
, we derive the zero-energy point equation of the energy band in parameter
space under periodic boundary conditions. The topological properties of the
system originate from this equation. In Sec. \ref{Topology originated from
nodal loop}, we focus on the case where spin-orbit coupling is negligible.
We theoretically investigate the pumping spin of every energy band using
perturbation theory. Furthermore, we present numerical results for the
general case of spin-orbit coupling, which demonstrate that the pumping spin
is a universal topological invariant for the system. In Sec. \ref%
{Edge-pumping-spin flip}, we introduce the concepts of spin-flip current and
pumping-spin flip. Our numerical simulations demonstrate that the
distribution of the pumping-spin flip exhibits marginal characteristics, and
the aggregate of the marginal pumping-spin flips is found to be quantized.
we draw conclusions in Sec. \ref{Summary}. Some detailed derivations are
given in the Appendix.

\section{Model and symmetry}

\label{Model and symmetry}

Considering one-dimensional RM model on $2N$ lattice with spin-orbit
coupling, the Hamiltonian is

\begin{equation}
H=\sum\limits_{\sigma =\uparrow ,\downarrow }H_{\sigma }+H_{\mathrm{so}},
\label{H}
\end{equation}%
where the coupling-free part is

\begin{eqnarray}
H_{\sigma } &=&\sum\limits_{j=1}^{2N}(-1)^{\sigma }[\frac{1+(-1)^{j}\delta }{%
2}\left( c_{j,\sigma }^{\dag }c_{j+1,\sigma }+\mathrm{H.c.}\right)  \notag \\
&&-\left( -1\right) ^{j}Vc_{j,\sigma }^{\dag }c_{j,\sigma }],
\end{eqnarray}%
and the spin-orbit coupling term is 
\begin{equation}
H_{\mathrm{so}}=\frac{\lambda }{2}\sum\limits_{j=1}^{2N}\sum\limits_{\sigma
=\uparrow ,\downarrow }[(-1)^{\sigma }c_{j,\sigma }^{\dag }c_{j+1,-\sigma }+%
\mathrm{H.c.}].
\end{equation}%
Here, $H_{\uparrow }+H_{\downarrow }$ can be seen as two independent RM
models with opposite spin. And $c_{j,\sigma }^{\dag }$ is fermion creation
operator at site $j$, $\sigma =\uparrow ,\downarrow $ denotes spin
polarization. We have $c_{2N+1,\sigma }=c_{1,\sigma }$\ for the system with
periodic boundary condition, while $c_{2N+1,\sigma }=0$\ for open boundary
condition. The strength of hopping is dimensionless and spin-dependent,
given by $(-1)^{\uparrow }=-(-1)^{\downarrow }=1$. Term $H_{\mathrm{so}}$
characterizes the spin-orbit coupling with the strength $\lambda /2$.

For the Hamiltonian $H_{\uparrow }+H_{\downarrow }$, both total fermion
number $n$\ with spin $\sigma =\uparrow ,\downarrow $\ and total spin
component $s_{z}$\ are conservative, that is%
\begin{equation}
\left[ n_{\sigma },H_{\uparrow }+H_{\downarrow }\right] =\left[
s_{z},H_{\uparrow }+H_{\downarrow }\right] =0,
\end{equation}%
where%
\begin{equation}
n_{\sigma }=\sum\limits_{j=1}^{2N}c_{j,\sigma }^{\dag }c_{j,\sigma },
\end{equation}%
and
\begin{equation}
s_{\alpha }=\sum\limits_{j=1}^{2N}s_{j}^{\alpha }.
\end{equation}%
Here the spin operators $s^{\alpha }$ $\left( \alpha =x,y,z\right) $\ at $j$%
th site (in units of $\hbar /2$) are defined as

\begin{equation}
s_{j}^{\alpha }=\left( 
\begin{array}{cc}
c_{j,\uparrow }^{\dag } & c_{j,\downarrow }^{\dag }%
\end{array}%
\right) \sigma ^{\alpha }\left( 
\begin{array}{c}
c_{j,\uparrow } \\ 
c_{j,\downarrow }%
\end{array}%
\right) ,
\end{equation}%
where $\sigma ^{\alpha }$ $\left( \alpha =x,y,z\right) $ are Pauli matrices,
given as

\begin{eqnarray}
\sigma ^{x} &=&\left( 
\begin{array}{cc}
0 & 1 \\ 
1 & 0%
\end{array}%
\right) ,\sigma ^{y}=\left( 
\begin{array}{cc}
0 & -i \\ 
i & 0%
\end{array}%
\right) ,  \notag \\
\sigma ^{z} &=&\left( 
\begin{array}{cc}
1 & 0 \\ 
0 & -1%
\end{array}%
\right) .
\end{eqnarray}%
We note that the term $H_{\mathrm{so}}$\ breaks the conservations of $%
n_{\sigma }$\ and $s_{z}$, but remains the conservation of total fermion
number $n=n_{\uparrow }+n_{\downarrow }$. In addition, defining a
spin-flipping operator $R$ as follows

\begin{equation}
Rc_{j,\sigma }R^{-1}=c_{j,-\sigma },
\end{equation}%
which flips a spin to the opposite direction, we have

\begin{equation}
RHR^{-1}=-H.
\end{equation}%
This means that the energy levels of $H$ are symmetric with the respect to $%
0 $.

\section{Topological pumping spin}

\label{Topological pumping spin}

In this work, we investigate the topology of the Hamiltonian $H$ by
considering the time-dependent parameter $\delta =\delta (t)$\ and $V=V(t)$.
To proceed, we first give a brief review of features of the Hamiltonian $%
H_{\sigma }$. It can be diagonalized in the form%
\begin{equation}
H_{\sigma }=\sum_{k}\varepsilon _{k,\sigma }(\alpha _{k,\sigma }^{\dag
}\alpha _{k,\sigma }-\beta _{k,\sigma }^{\dag }\beta _{k,\sigma }),
\end{equation}%
where the spectrum is given by 
\begin{equation}
\varepsilon _{k,\sigma }=\sqrt{V^{2}+|\gamma _{k}|^{2}}.
\end{equation}%
Here $\gamma _{k}=\left[ (1-\delta )+(1+\delta )e^{ik}\right] /2$\ and
wavevector $k=2n\pi /N$, $n=1,2,...,N$.\ Two sets of fermion operators $%
\left\{ \alpha _{k,\sigma },\beta _{k,\sigma }\right\} $\ can be extracted
from the single-particle solution of the equation, given by%
\begin{equation}
H_{\sigma }|\psi _{k,\sigma }^{\pm }\rangle =\pm \varepsilon _{k,\sigma
}|\psi _{k,\sigma }^{\pm }\rangle ,
\end{equation}%
where $|\psi _{k,\sigma }^{+}\rangle =\alpha _{k,\sigma }^{\dag }|0\rangle $
and $|\psi _{k,\sigma }^{-}\rangle =\beta _{k,\sigma }^{\dag }|0\rangle $.
Here $|0\rangle $ is the vacuum state satisfying $\alpha _{k,\sigma
}|0\rangle =\beta _{k,\sigma }|0\rangle =0.$ For the adiabatic time evolution
along an arbitrary loop with 
\begin{equation}
\delta (t+T)=\delta (t),V(t+T)=V(t),
\end{equation}%
the pumping charge $Q_{\sigma }=\int_{0}^{T}J_{\sigma }\mathrm{d}t$ is shown
to be a demonstration of the Chern number, where the current across two
neighboring sites is equivalent to the $k$ integral of the Berry curvature,
that is%
\begin{equation}
J_{\sigma }=\frac{i}{2\pi }\int_{0}^{2\pi }[(\partial _{t}\langle \psi
_{k,\sigma }^{-}|)\partial _{k}|\psi _{k,\sigma }^{-}\rangle -(\partial
_{k}\langle \psi _{k,\sigma }^{-}|)\partial _{t}|\psi _{k,\sigma
}^{-}\rangle ]\mathrm{d}k.
\end{equation}%
We have the following topological characteristics%
\begin{equation}
Q_{\sigma }=\left\{ 
\begin{array}{cc}
(-1)^{\sigma }, & \text{enclosed }\left( 0,0\right)  \\ 
0, & \text{otherwise}%
\end{array}%
\right. ,
\end{equation}%
which arises from the degenerate point at $\left( \delta ,V\right) =\left(
0,0\right) $. The pumping charge can be computed by quasi-adiabatic passage.
Obviously, for the ground state of $H_{\uparrow }+H_{\downarrow }$ the
obtained pumping charge $Q_{\uparrow }+Q_{\downarrow }$\ is vanishing no
matter the loop in the $\delta -V$ plane encloses the origin or not. Then
the topological invariant in $H_{\uparrow }+H_{\downarrow }$\ cannot be
detected via the pumping charge.

Nevertheless, it is obvious that the sub-Hamiltonians $H_{\uparrow }$\ and $%
H_{\downarrow }$\ are really topologically non-trivial. We need an
alternative topological invariant to characterize such topology.
Techniquely, an amount of pumping charge $Q_{\sigma }$\ is always associated
with an amount of pumping spin $\left( -1\right) ^{\sigma }Q_{\sigma }$.
Then we can employ the total pumping spin for the ground state, defined as $%
S=Q_{\uparrow }-Q_{\downarrow }$, as topological invariant, which obeys%
\begin{equation}
S=\left\{ 
\begin{array}{cc}
2, & \text{enclosed }\left( 0,0\right) \\ 
0, & \text{otherwise}%
\end{array}%
\right. .
\end{equation}%
In the following, we will investigate what happens in the presence of
spin-orbit coupling.

\section{Nodal circle}

\label{Nodal circle}

In this section, we focus on the solution of the system in the presence of
spin-orbit coupling. Under periodic boundary conditions, the system has
translational symmetry, with each unit cell including four degree of
freedom. Employing Fourier transformation 
\begin{equation}
\left( 
\begin{array}{c}
c_{2j-1,\uparrow } \\ 
c_{2j,\uparrow } \\ 
c_{2j-1,\downarrow } \\ 
c_{2j,\downarrow }%
\end{array}%
\right) =\frac{e^{ikj}}{\sqrt{N}}\left( 
\begin{array}{c}
a_{k,\uparrow } \\ 
b_{k,\uparrow } \\ 
b_{k,\downarrow } \\ 
a_{k,\downarrow }%
\end{array}%
\right) ,
\end{equation}%
the Hamiltonian can be written as%
\begin{equation}
H=\sum_{2\pi >k>0}\Psi _{k}^{\dag }h_{k}\Psi _{k},
\end{equation}%
where the operator vector is defined as

\begin{equation}
\Psi _{k}^{\dag }=\left( 
\begin{array}{cccc}
a_{k,\uparrow }^{\dag } & b_{k,\uparrow }^{\dag } & a_{k,\downarrow }^{\dag }
& b_{k,\downarrow }^{\dag }%
\end{array}%
\right) ,
\end{equation}%
and the core matrix is

\begin{equation}
h_{k}=\left( 
\begin{array}{cccc}
V & \zeta & \xi & 0 \\ 
\zeta ^{\ast } & -V & 0 & -\xi ^{\ast } \\ 
\xi ^{\ast } & 0 & V & -\zeta ^{\ast } \\ 
0 & -\xi & -\zeta & -V%
\end{array}%
\right).
\end{equation}%
Here the $k$-dependent factors are given by

\begin{eqnarray}
\zeta &=&\left[ 1-\delta +\left( 1+\delta \right) e^{-ik}\right] /2,  \notag
\\
\xi &=&\lambda \left( 1-e^{-ik}\right) /2.  \label{variable}
\end{eqnarray}%
The four eigenvalues of $h_{k}$ can be expressed as

\begin{equation}
\varepsilon _{\mu ,\nu }=\mu \sqrt{\zeta \zeta ^{\ast }+\xi \xi ^{\ast
}+V^{2}+\nu \sqrt{4V^{2}\xi \xi ^{\ast }+\left( \zeta \xi ^{\ast }+\xi \zeta
^{\ast }\right) ^{2}}},  \label{spectrum}
\end{equation}%
with $\mu $, $\nu =\pm $.\ The derivations in the appendix show that the
energy gap between positive and negative bands closes only when $k=0$ or $%
\pi $. The corresponding zero-energy points form a nodal loop in the $\delta
-V$ plane, obeying the equation 
\begin{equation}
\delta ^{2}+V^{2}=\lambda ^{2}.  \label{nodal circle}
\end{equation}%
Obviously, it is a circle with radius $\lambda $ centered at the origin.\ In
Fig. \ref{fig1}, the spectral structure in the parameter space, obtained
from the spectrum $\varepsilon _{\mu ,\nu }$, is plotted for different
values of $\lambda $ as an illustration.

\begin{figure*}[tbh]
\centering
\includegraphics[width=1\textwidth]{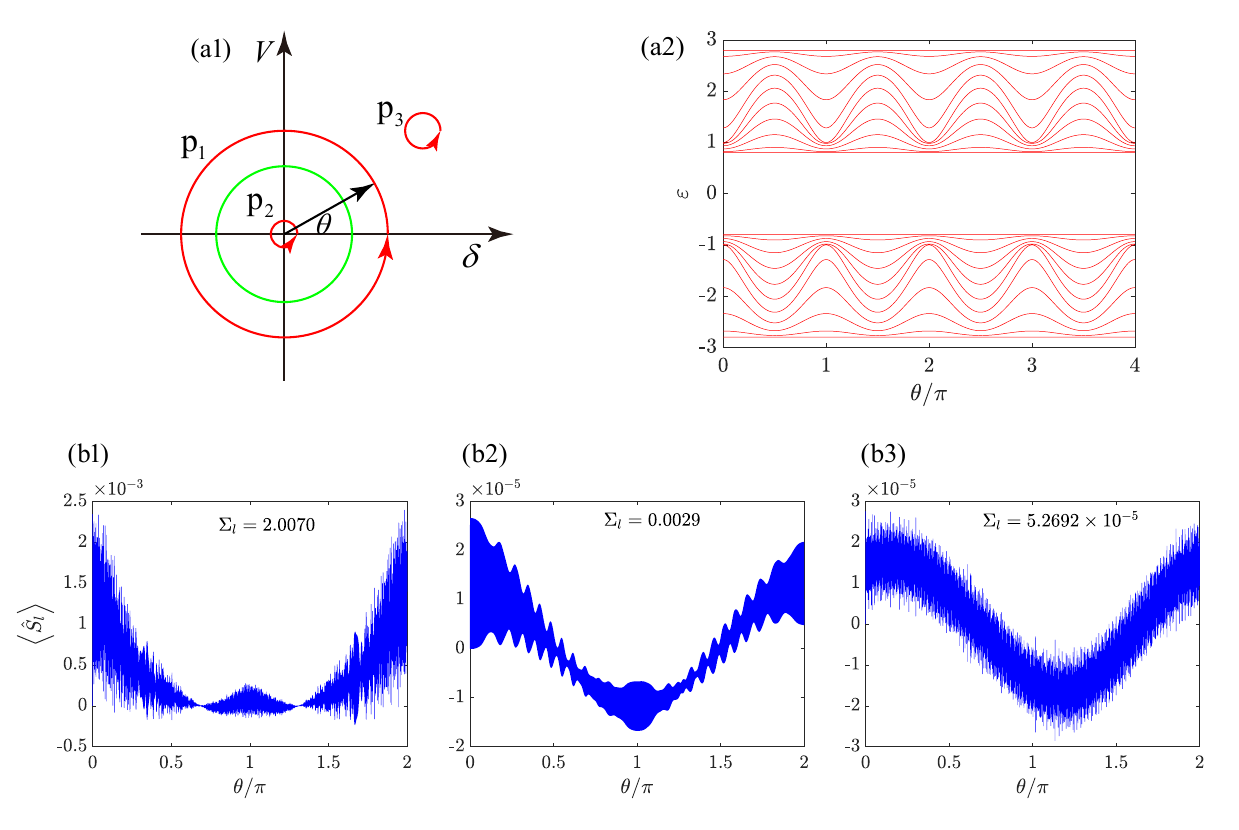}
\caption{The schematics of the adiabatic passages, the corresponding energy
spectrum, pumping spin currents, and pumping spins of the ground state for
several quasi-adiabatic passages. The driven system adopts periodic boundary
conditions to simulate the dynamic behavior of the topology originating from
the nodal circles. (a1) The green circle represents the nodal circle for $%
\protect\lambda =1$ in the $\protect\delta -V$ plane, while the red loops $%
\mathrm{p}_{1}$, $\mathrm{p}_{2}$ and $\mathrm{p}_{3}$ are three adiabatic
passages in parameter space with an anticlockwise direction. The
corresponding parameter equations are as follows: $\mathrm{p}_{1}$: $\protect%
\delta =1.8\cos \left( \protect\omega t\right) $, $V=1.8\sin \left( \protect%
\omega t\right) $; $\mathrm{p}_{2}$: $\protect\delta =0.05\cos \left( 
\protect\omega t\right) $, $V=0.05\sin \left( \protect\omega t\right) $; $%
\mathrm{p}_{3}$: $\protect\delta =0.1\cos \left( \protect\omega t\right) +2$%
, $V=0.1\sin \left( \protect\omega t\right) +2$. (a2) shows the plot of the
instantaneous spectrum of the time-dependent Hamiltonian for the passage $%
\mathrm{p}_{1}$, where $\protect\theta =\protect\omega t$ and the
evolutionary period of every energy level is $2\protect\pi $. (b1), (b2),
and (b3) correspond to the spin currents $\left\langle \hat{S}%
_{l}\right\rangle=\left\langle \protect\phi \left( t\right) \right\vert \hat{%
S}_{l}\left\vert \protect\phi \left( t\right) \right\rangle $ and pumping
spin $\Sigma _{l}$\ for the passages $\mathrm{p}_{1}$, $\mathrm{p}_{2}$ and $%
\mathrm{p}_{3}$, respectively, obtained from the numerical results for Eq. (%
\protect\ref{spincurrent}) and Eq. (\protect\ref{pumpingspin}). Here the
position $l$\ is taken arbitrarily due to the translational symmetry. The
results indicate that pumping spin is nearly quantized, being $2$ or $0$,
determined by whether the adiabatic passage encircles the nodal circle or
not. Other parameters are $N=10$, $\protect\omega =0.001$, and $\Delta t=%
\protect\pi /200$. }
\label{fig2}
\end{figure*}

\begin{figure*}[tbh]
\centering
\includegraphics[width=1\textwidth]{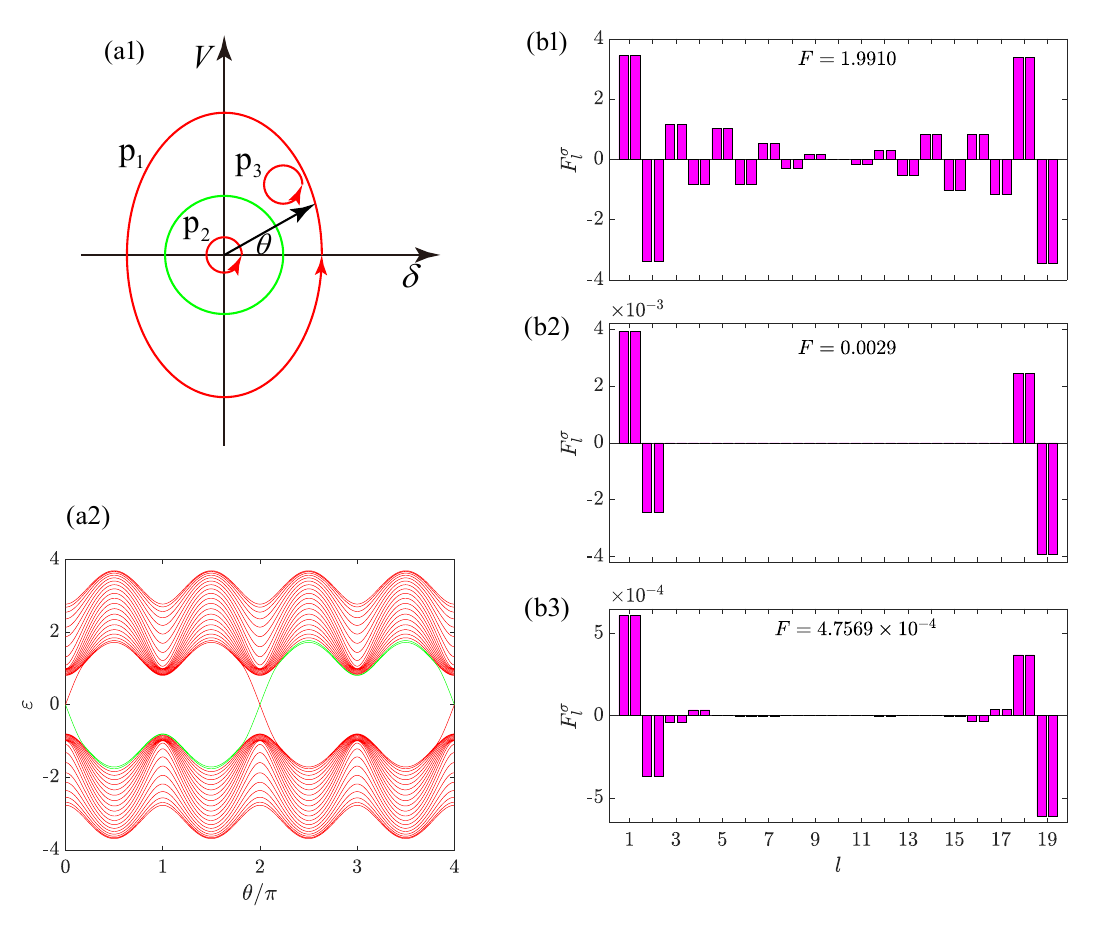}
\caption{The schematics of the adiabatic passages, the corresponding energy
spectrum, pumping-spin flip distribution, and total edge-pumping-spin flip
of the ground state for several quasi-adiabatic passages. The driven system
adopts open boundary conditions to simulate the dynamic behavior of the
boundary-bulk correspondence. (a1) The green circle represents the nodal
circle for $\protect\lambda =1$ in the $\protect\delta -V$ plane, while the
red loops $\mathrm{p}_{1}$, $\mathrm{p}_{2}$ and $\mathrm{p}_{3}$ are three
adiabatic passages in parameter space with an anticlockwise direction. The
corresponding parameter equations are as follows: $\mathrm{p}_{1}$: $\protect%
\delta =1.8\cos \left( \protect\omega t+\protect\theta _{0}\right) $, $%
V=2.7\sin \left( \protect\omega t+\protect\theta _{0}\right) $; $\mathrm{p}%
_{2}$: $\protect\delta =0.05\cos \left( \protect\omega t+\protect\theta %
_{0}\right) $, $V=0.05\sin \left( \protect\omega t+\protect\theta %
_{0}\right) $; $\mathrm{p}_{3}$: $\protect\delta =0.1\cos \left( \protect%
\omega t+\protect\theta _{0}\right) +1.3$, $V=0.1\sin \left( \protect\omega t+\protect\theta _{0}\right) +1.3$. (a2)
shows the plot of the instantaneous spectrum of the time-dependent
Hamiltonian for the passage $\mathrm{p}_{1}$, where $\protect\theta =\protect%
\omega t+\protect\theta _{0}$ and the evolutionary period of every energy
level is $4\protect\pi $. There is a crossing point between two energy
levels of $2$-fold degeneracy edge states at $\protect\theta =2\protect\pi $%
. One of the edge energy levels is indicated in green. It indicates that any
edge states completes its adiabatic cycle with a period\ of $4\protect\pi $.
(b1), (b2), and (b3) correspond to the distribution of the pumping-spin
flip\ $F_{l}^{\protect\sigma }$\ and total edge-pumping-spin $F$\ for the
passages $\mathrm{p}_{1}$, $\mathrm{p}_{2}$ and $\mathrm{p}_{3}$,
respectively, obtained from the numerical results for Eq. (\protect\ref{flip}%
) and Eq. (\protect\ref{Fto}). We observe that the pumping-spin flip\ $%
F_{l}^{\protect\sigma }$\ vanishes in the bulk region. The results indicate
that total edge-pumping-spin flip $F$\ is nearly quantized, being $2$ or $0$%
, determined by whether the adiabatic passage encircles the nodal circle or
not. Other parameters are $N=10$, $\protect\omega =0.0005$, $\protect\theta %
_{0}=\protect\pi /2$ and $\Delta t=\protect\pi /200$. }
\label{fig3}
\end{figure*}

\section{Topology originated from nodal loop}

\label{Topology originated from nodal loop}

It has been shown that the quantized Thouless pumping of charge and spin in
the RM model, in the absence of spin-orbit coupling, originates from the
degeneracy point at the origin. Furthermore, this is also true when multiple
degenerate points are involved. Now, a natural question arises: what happens
when these degenerate points form a degenerate loop, which contains an
infinite number of degenerate points? Our strategy consists of two steps:
(i) We perform an analytical analysis by considering an adiabatic loop that
is far from the nodal circle; (ii) We conduct numerical simulations to
verify the results obtained.

Now we focus on the first step. The spin-orbit coupling term $H_{\mathrm{so}%
} $\ can be regarded as perturbation under the condition $\delta
^{2}+V^{2}\gg \lambda ^{2}$. In the following, we only focus on the negative
energy bands. Applying perturbation method, we can get zero-order
approximation eignstates of $h_{k}$ for two negative energy bands, given by%
\begin{equation}
\left\vert \varphi _{k}^{\pm }\right\rangle =\frac{|\psi _{k,\downarrow
}^{-}\rangle \mp \mathrm{sgn}\left( \lambda \right) e^{i\Lambda _{k}}|\psi
_{k,\uparrow }^{-}\rangle }{\sqrt{2}},
\end{equation}%
where the factor $\Lambda _{k}$\ is given by%
\begin{eqnarray}
\Lambda _{k} &=&\arg \left[ \sin \left( \phi _{k}-\frac{k}{2}\right) +i\cos
\theta _{k}\cos \left( \phi _{k}-\frac{k}{2}\right) \right]  \notag \\
&&-\phi _{k},
\end{eqnarray}%
with 
\begin{eqnarray}
\theta _{k} &=&\arctan \left[ \frac{\sqrt{1+\delta ^{2}+\left( 1-\delta
^{2}\right) \cos k}}{\sqrt{2}V}\right] ,  \notag \\
\phi _{k} &=&\arctan \left[ \frac{\left( 1+\delta \right) \sin k}{\left(
1-\delta \right) +\left( 1+\delta \right) \cos k}\right] .
\end{eqnarray}

Accordingly, for an adiabatic time evolution along an arbitrary loop far
from the nodal loop, the pumping charge for each band is $Q^{\pm
}=\int_{0}^{T}J^{\pm }\mathrm{d}t$. Here the current across two neighboring
sites is%
\begin{equation}
J^{\pm }=\frac{i}{2\pi }\int_{0}^{2\pi }[(\partial _{t}\langle \varphi
_{k}^{\pm }|)\partial _{k}|\varphi _{k}^{\pm }\rangle -(\partial _{k}\langle
\varphi _{k}^{\pm }|)\partial _{t}|\varphi _{k}^{\pm }\rangle ]\mathrm{d}k.
\end{equation}%
Since the factor $\Lambda _{k}$ has no contribution to the Berry curvature,
direct derivations show that%
\begin{equation}
J^{\pm }=\frac{1}{2}\left( J_{\uparrow }+J_{\downarrow }\right) .
\end{equation}%
Then the pumping charge $Q^{\pm }=\left( Q_{\uparrow }+Q_{\downarrow
}\right) /2$\ is vanishing no matter the loop in the $\delta -V$\ plane
encloses the nodal circle or not. In contrast, the pumping spin for the
ground state can also be regarded as topological invariant, obeying%
\begin{equation}
S=\left\{ 
\begin{array}{cc}
2, & \text{enclosed the nodal circle} \\ 
0, & \text{otherwise}%
\end{array}%
\right. .
\end{equation}%
This conclusion should still hold true when the adiabatic loop approaches to
the nodal circle due to the fact that the Chern number must be an integer.

Now we turn to the second step. In order to verify and demonstrate the
conclusions, numerical simulations are performed for quasi-adiabatic time
evolution in finite systems. The driven system adopts periodic boundary
conditions to simulate the dynamic behavior of the topology originating from
the nodal circles. To this end, we introduce the local charge current $\hat{J%
}_{l,\sigma }$ and spin current $\hat{S}_{l}$ operators as%
\begin{eqnarray}
\hat{J}_{l,\sigma } &=&\frac{1+\left( -1\right) ^{l}\delta }{2i}(-1)^{\sigma
}c_{l,\sigma }^{\dag }c_{l+1,\sigma }+\mathrm{H.c.,} \\
\hat{S}_{l} &=&\frac{1+\left( -1\right) ^{l}\delta }{2i}\sum\limits_{\sigma
}c_{l,\sigma }^{\dag }c_{l+1,\sigma }+\mathrm{H.c.,}  \label{spincurrent}
\end{eqnarray}%
for the dimer across two neighboring sites $l$\ and $l+1$ with $l\in \left[
1,2N\right] $.

In the following we are going to calculate the pumping charge and spin of
ground state for a quasi-adiabatic passage in the parameter space. We
consider the time-dependent Hamiltonian $H\left( t\right) $ with
periodically varying parameters as follows, 
\begin{eqnarray}
\delta \left( t\right) &=&R_{1}\cos \left( \omega t+\theta _{0}\right)
+\delta _{0},  \notag \\
V\left( t\right) &=&R_{2}\sin \left( \omega t+\theta _{0}\right) +V_{0},
\label{parameter}
\end{eqnarray}%
which is a closed loop with a center at $\left( \delta _{0},V_{0}\right) $
in the $\delta -V$ plane. Here, $\omega $ controls the varying speed of the
Hamiltonian and the period is $T=2\pi /\omega $. The pumping charge and spin
for an evolution period can be expressed as 
\begin{eqnarray}
Q_{l,\sigma } &=&\int_{0}^{T}\left\langle \phi \left( t\right) \right\vert 
\hat{J}_{l,\sigma }\left\vert \phi \left( t\right) \right\rangle \text{%
\textrm{d}}t, \\
\Sigma _{l} &=&\int_{0}^{T}\left\langle \phi \left( t\right) \right\vert 
\hat{S}_{l}\left\vert \phi \left( t\right) \right\rangle \text{\textrm{d}}t,
\end{eqnarray}
where $\left\vert \phi \left( t\right) \right\rangle $ is the evolved state
from the initial ground state $\left\vert \phi \left( 0\right) \right\rangle 
$ of $H\left( 0\right) $. The practical computation is performed by using a
uniform mesh in time discretization. Time is discretized into $t_{m}$, with $%
t_{0}=0$ and $t_{M}=T$. For a given initial eigenstate $\left\vert \phi
\left( 0\right) \right\rangle $, the time-evolved state is computed using 

\begin{equation}
\left\vert \phi \left( t_{n}\right) \right\rangle =\mathcal{T}%
\prod\limits_{m=1}^{n}\exp \left[ -iH\left( t_{m-1}\right) \left(
t_{m}-t_{m-1}\right) \right] \left\vert \phi \left( 0\right) \right\rangle ,
\end{equation}
where $\mathcal{T}$ is the time-order operator. In the simulation, the value
of $M$ is considered sufficiently large to obtain a convergent result and
the pumping charge and spin is

\begin{eqnarray}
Q_{l,\sigma } &\approx &\sum_{m=1}^{M}\left\langle \phi \left( t_{m}\right)
\right\vert \hat{J}_{l,\sigma }\left( t_{m}\right)\left\vert \phi \left( t_{m}\right)
\right\rangle \Delta t, \\
\Sigma _{l} &\approx &\sum_{m=1}^{M}\left\langle \phi \left( t_{m}\right)
\right\vert \hat{S}_{l}\left( t_{m}\right)\left\vert \phi \left( t_{m}\right) \right\rangle
\Delta t,  \label{pumpingspin}
\end{eqnarray}%
where, $\Delta t=\left( t_{l}-t_{l-1}\right) $ is the time step. In the
process of computation, the value of $\omega $ should be sufficiently small
to fulfill the requirement of quasi-adiabatic evolution.

In Fig. \ref{fig2}, the schematics of the adiabatic passages, the
corresponding energy spectrum, pumping spin currents, and pumping spins of
the ground state for several quasi-adiabatic passages are presented. The
results indicate that the pumping spin is nearly quantized, being $2$ or $0$%
, determined by whether the adiabatic passage encircles the nodal circle or
not. The results are similar to the pumping charge of the ground state of a
RM model with single degeneracy point, where the pumping charge is $\pm 1$
or $0$, determined by whether the adiabatic loop encircles the degenerate
point or not. This strongly implies that the topology of the spinful RM
model with spin-orbit coupling is same as that in single spinless RM model
with a degenerate point.

\section{Edge-pumping-spin flip}

\label{Edge-pumping-spin flip}

In the previous section, we explored the behavior of pumped spin under
periodic boundary conditions. An intriguing question arises when considering
open boundaries: what changes occur? In the following, we focus on the
system with open boundary conditions. In this situation, we always have $%
Q_{\uparrow }=Q_{\downarrow }=0$ for any adiabatic passage due to the
disconnection at the boundary. It is obvious that the total pumping spin is
always zero for any adiabatic passage and thus cannot be used to
characterize the topological feature.\ Nevertheless, we note that there
should be an accumulation of particles at the ends of the chain, which may
induce a spin-flip current. The operator of the spin-flip current can be
defined as follows

\begin{equation}
\mathcal{J}_{l}^{\sigma }=\frac{\lambda }{2i}c_{l,\sigma }^{\dag
}c_{l+1,-\sigma }+\mathrm{H.c.,}
\end{equation}%
where $l\in \left[ 1,2N-1\right] $. In comparison with the normal current
operator, it measures the current across sites $l$\ and $l+1$, associated
with spin flipping. When there is no spin-orbit coupling, it becomes clear
that no spin-flip current can flow, as the channels remain disconnected.
However, in the presence of spin-orbit coupling, a transport channel between
two opposite spins exists, potentially creating an accumulation of spin
flips. Similarly, we can define the pumped spin-flip as

\begin{equation}
F_{l}^{\sigma }=\int_{0}^{2T}\left\langle \phi \left( t\right) \right\vert 
\mathcal{J}_{l}^{\sigma }\left\vert \phi \left( t\right) \right\rangle dt,
\end{equation}%
for a given adiabatic passage. Here, $T$ is the period of the time-dependent
Hamiltonian, with parameters as described in Eq. (\ref{parameter}).\ In
general, the integral period is $T$, corresponding to a single cycle of the
quasi-adiabatic time evolution along the adiabatic passage. However, for an
open chain, a double cycle of the quasi-adiabatic time evolution along the
adiabatic passage should be considered to ensure that every initial
single-particle eigenstate can evolve back. To demonstrate this point, we
plot the energy level structure for the passage in Fig. \ref{fig3} (a2). We
can observe that the edge states do not revert after a single adiabatic
passage cycle, whereas all other eigenstates do.

To perform numerical simulation of the quantity, we have 
\begin{equation}
F_{l}^{\sigma }\approx \sum_{m=1}^{2M}\left\langle \phi \left( t_{m}\right)
\right\vert \mathcal{J}_{l}^{\sigma }\left( t_{m}\right) \left\vert \phi \left( t_{m}\right)
\right\rangle \Delta t,  \label{flip}
\end{equation}%
which is similar to the formulation presented in Eq. (\ref{pumpingspin}).
The details and the results are elucidated in Fig. \ref{fig3}. It displays
several adiabatic loops in parameter space, while (a2) shows the energy
spectrum of the adiabatic loop encircling the nodal circle in (a1). We find
that two distinct energy levels, which are separated from the rest,
correspond to the energies of the edge states. These two energy levels cross
once after one period $T$, causing the edge state adiabatically evolve along
the green energy level depicted in (a2), and then return to its initial
state after evolving for two periods. In (b1), (b2) and (b3) we present the
corresponding simulation results for the pumping-spin flip defined in Eq. (%
\ref{flip}). The results in (b1) demonstrate that $F_{l}^{\sigma
}=F_{l}^{-\sigma }$, indicating that the pumping-spin flip is
position-dependent. Specifically, the magnitude of the pumping-spin flip
decreases from the edges to the middle of the open chain, reaching zero at
the center region. Furthermore, the pumping-spin flip exhibits symmetry with
respect to the center of the chain, but with opposite signs. To characterize
the topology of edge state, we introduce the concept of total pumping-spin
flip for a half chain, defined as follows 
\begin{equation}
F=\sum\limits_{l=1,\sigma }^{N-1}F_{l}^{\sigma },  \label{Fto}
\end{equation}%
which is the sum of the pumping-spin flip over half of the sites. The
numerical results reveal that the total pumping-spin flip approaches the
value of $2$ with high precision when the adiabatic loop encircles the nodal
circle. Conversely, when the adiabatic loop is either entirely within or
outside the nodal circle, the total pumping-spin flip is close to $0$ with
minimal deviation. Hence, the quantized total pumping-spin flip can serve as
an indicator for the topology of the system under open boundary conditions.
This result reveals the dynamic behavior associated with the boundary-bulk
correspondence.

\section{Summary}

\label{Summary}

In summary, we investigate the dynamic demonstration on topology in
association with nodal loop in a system with periodic and open boundary
conditions. In contrast to the previous works, the nodal loop lies in a $2$D
parameter space, rather than $3$D space. The topology studied in the present
work originates from a degenerate line, while in the previous work, such as
that in Ref. \cite{WR1}, the related topology feature essentially originates
from isolated degenerate point. Furthermore, we also investigate a dynamic
manifestation of the boundary-bulk correspondence by computing the
pumping-spin flip for different adiabatic passages under the open boundary
conditions. The results indicate that pumping-spin flips at the edges are
also the quantized only for double periods of a closed passage. Our findings
not only propose a new origin for the topology but also provide another
method for dynamically detecting the boundary-bulk correspondence.

\acknowledgments This work was supported by the National Natural Science
Foundation of China (under Grant No. 12374461).

\section*{Appendix}

\appendix\setcounter{equation}{0} \renewcommand{\theequation}{A%
\arabic{equation}}

In this appendix, we derive the equation for the nodal loop based on the
spectrum provided in Eq. (\ref{spectrum}). When the energy gap closes, we
obtain the following result

\begin{equation}
\left( \zeta \zeta ^{\ast }+\xi \xi ^{\ast }+V^{2}\right) ^{2}=4V^{2}\xi \xi
^{\ast }+\left( \zeta \xi ^{\ast }+\xi \zeta ^{\ast }\right) ^{2},
\end{equation}%
which implies that

\begin{equation}
\left( \zeta \zeta ^{\ast }-\xi \xi ^{\ast }+V^{2}\right) ^{2}=\left( \zeta
\xi ^{\ast }-\zeta ^{\ast }\xi \right) ^{2}.
\end{equation}%
Furthermore, by applying Eq. (\ref{variable}), we find that

\begin{equation}
\zeta \xi ^{\ast }=\frac{\lambda }{2}\left[ \delta \left( \cos k-1\right)
-i\sin k\right] ,
\end{equation}%
and then 
\begin{equation}
\left( \zeta \zeta ^{\ast }-\xi \xi ^{\ast }+V^{2}\right) ^{2}=-\lambda
^{2}\sin ^{2}k.
\end{equation}%
Clearly, the equation has solutions only when $k$ is either $0$ or $\pi $.
For $k=0$, we get

\begin{equation}
\zeta \zeta ^{\ast }-\xi \xi ^{\ast }+V^{2}=1+V^{2}=0,
\end{equation}%
which indicates that there is no solution. For $k=\pi $, we get the solution

\begin{equation}
\delta ^{2}+V^{2}=\lambda ^{2},
\end{equation}%
which represents a circle with radius $\lambda $.

\end{document}